\documentclass[twocolumn,aps,prl,showpacs]{revtex4}
\usepackage{graphics}
\begin{document}

\title{Self-diffusion in sheared colloidal suspensions: violation of 
fluctuation-dissipation relation}
\author{Grzegorz Szamel}
\affiliation{Department of Chemistry, 
Colorado State University, Fort Collins, CO 80525}

\date{\today}


\begin{abstract}
Using memory-function formalism we show that in sheared colloidal
suspensions the fluctuation-dissipation theorem for self-diffusion, 
\textit{i.e} Einstein's relation between self-diffusion and mobility
tensors, is violated and propose a new way to measure this violation
in Brownian Dynamics simulations. We derive mode-coupling expressions
for the tagged particle friction tensor and for an effective,
shear-rate dependent temperature. 

\end{abstract}
\maketitle

There has been a lot of interest recently in non-equilibrium
behavior of colloidal systems \cite{Faraday}. On the practical side,
it has been stimulated by the importance of non-equilibrium
properties for the preparation and processing of colloidal materials.
From the more fundamental perspective, colloidal suspensions serve as
model soft-glassy systems: they have properties similar to those
of more complex soft materials but are simple enough to
allow for detailed microscopic, experimental and theoretical 
investigations. Additional impetus % for studies of non-equilibrium behavior 
came from analysis of simple statistical mechanical models 
(\textit{i.e.} fully connected spin systems) that predicted 
violation of the fluctuation-dissipation theorem (FDT) out of
equilibrium \cite{CK}, 
and an intriguing connection between non-equilibrium and glassy
properties \cite{neqglassy}.
Subsequently, FDT violation was found by means of computer simulations
in model supercooled fluids \cite{BK,BB}.

Although non-equilibrium phenomenology of colloidal suspensions 
is reasonably well described by so-called ``schematic models'' \cite{BBK,FC2}, 
it is of great fundamental and practical
interest to develop more microscopic approaches. First, connection
between non-equilibrium and glassy behavior, if it also exists 
for colloidal systems, may provide new insights into the glass
transition problem. Second, simulational and experimental studies of
colloidal systems provide detailed information that cannot be described
in terms of schematic models. Third, a more microscopic
approach would allow one to correlate microscopic properties  
and macroscopic behavior. 

Recent investigations of the non-equilibrium behavior can be divided into
two categories. Initially, transient behavior of glassy
systems, \textit{i.e.} aging, attracted the most attention
\cite{CK,neqglassy,BK}. 
Microscopic, theoretical analysis of aging is in its infancy \cite{Latz}.
More recently, relaxation under steady shear 
was investigated \cite{BB,Onuki,BBK}. 
Analysis of sheared suspensions is easier: 
stationary nature of the shear flow restores 
time-translational invariance and thus simplifies the problem
both conceptually and technically. 

Colloidal suspensions under steady shear flow were the subject of two
recent theoretical investigations \cite{FC,MR}. Both approaches were 
based on the ``least inadequate'' \cite{CHFH} 
microscopic theory of colloidal dynamics: mode-coupling theory (MCT)
\cite{Cates}.
The difference between them parallels the difference between
derivations of MCT for non-sheared colloids: Ref. \cite{FC}
used the projection operator method whereas Ref. \cite{MR} started from
generalized fluctuating hydrodynamics. Both approaches recovered 
the most important features of the soft-glassy rheology: 
accelerated relaxation of sheared colloidal fluids   
and shear melting of the colloidal glass. 
These phenomena were attributed to flow-induced advection of
density fluctuations and the resulting perturbation of the
``cage effect''.  
Neither work, however, addressed FDT violation.

The goal of this Letter is to investigate the origin of FDT violation 
for the simplest possible process: self-diffusion (\textit{i.e.} diffusion
of a tagged particle) in a sheared colloidal suspension \cite{IR,SBL}. 
In this case FDT amounts to Einstein's relation between
the self-diffusion tensor and the tagged particle mobility tensor.
We generalize the conventional memory function description
of self-diffusion to colloidal systems under shear and derive
a Green-Kubo-like relation for the self-diffusion tensor
in a sheared suspension. Next, we derive a Green-Kubo-like relation
for the tagged particle mobility tensor. We show that FDT violation
is associated with the non-equilibrium nature of the stationary,
shear-rate dependent probability distribution. We propose a new approach
to monitor FDT violation in Brownian Dynamics simulations that does
not require introducing an external perturbation. Finally, we use 
the memory function approach to derive MCT expressions for the
tagged particle friction tensor and an effective temperature. 

We start with the definition of the self-intermediate scattering function, 
$F_s(\mathbf{k}_1;\mathbf{k}_2;t)$: 
\begin{equation}\label{Fsdef}
F_s(\mathbf{k}_1;\mathbf{k}_2;t) = 
\left<n_s(\mathbf{k}_1) \exp(\Omega t) n_s(-\mathbf{k}_2)\right>.
\end{equation}
Here 
$n_s(\mathbf{k}_1)$ is the Fourier transform of the microscopic
density of particle number 1, \textit{i.e.} the tagged particle,
$n_s(\mathbf{k}_1) = e^{-i\mathbf{k}_1\cdot\mathbf{r}_1}$. Furthermore,
$\Omega$ is the $N$-particle evolution operator,
\textit{i.e.} the Smoluchowski operator \cite{commentHI2},
\begin{equation}\label{Sm}
\Omega = - D_0 \sum_l \frac{\partial}{\partial\mathbf{r}_l}\cdot \left(
- \frac{\partial}{\partial\mathbf{r}_l} + \beta\mathbf{F}_l
+\mathbf{v}(\mathbf{r}_l)
\right),
\end{equation}
where $D_0$ is the diffusion coefficient of an isolated colloidal 
particle, $\beta=1/(k_B T)$, $\mathbf{F}_l$ is a force acting
on the $l$th particle, and 
$\mathbf{v}(\mathbf{r)} = \mathbf{\Gamma}\cdot\mathbf{r}$ 
is the shear flow with $\mathbf{\Gamma} = 
\dot{\gamma} \hat{\mathbf{x}} \hat{\mathbf{y}}$ being the velocity gradient 
tensor ($\dot{\gamma}$ denotes the shear rate).
Finally, in Eq. (\ref{Fsdef}) $\left<\dots\right>$ denotes
the stationary, shear-rate dependent ensemble average at temperature $T$.
The probability distribution
stands to the right of the quantity being averaged and all operators
act on it as well as on everything else. 
Note that %the 
translational symmetry leads to %results in 
$F_s(\mathbf{k}_1;\mathbf{k}_2;t) \propto 
\delta_{\mathbf{k}_1(t),\mathbf{k}_2}$ where 
$\mathbf{k}_1(t) = \mathbf{k}_1 + ^t\!\mathbf{\Gamma}\cdot\mathbf{k}_1t$
with $^t\mathbf{\Gamma}$ being the transpose of $\mathbf{\Gamma}$.

To derive the memory function representation we start from the exact
expression for the Laplace transform ($LT$) of the time derivative 
of $F_s(\mathbf{k}_1;\mathbf{k}_2;t)$,
\begin{equation}\label{Fsktder}\nonumber
LT(\dot{F}_s) 
=\left< n_s(\mathbf{k}_1) \Omega \frac{1}{z-\Omega} n_s(-\mathbf{k}_2)
\right>.
\end{equation}
We define the projection operator 
on the space spanned by the tagged particle density,
\begin{eqnarray}\label{P}
\hat{P}_s &=& \sum_{\mathbf{q}} \dots
n_s(-\mathbf{q})\left>\right<n_s(\mathbf{q})\dots .
\end{eqnarray}
Note that, in contrast to Ref. \cite{FC}, 
the definition of $\hat{P}$ involves the 
stationary, shear-rate dependent distribution. 
Next, we follow the usual projection operator manipulations and arrive
at the following memory function representation of the time derivative
of $F_s(\mathbf{k}_1;\mathbf{k}_2;t)$,
\begin{eqnarray}\label{Fsktder2}\nonumber
LT(\dot{F}_s) &=& 
\sum_{\mathbf{k}_3} \left<n_s(\mathbf{k}_1)\Omega n_s(\mathbf{k}_3)\right> 
F_s(\mathbf{k}_3,\mathbf{k}_2;z)
\\ \nonumber &+& \sum_{\mathbf{k}_3}
\left<n_s(\mathbf{k}_1)\Omega\hat{Q}_s \frac{1}{z-\hat{Q}_s\Omega\hat{Q}_s}
\hat{Q}_s\Omega n_s(\mathbf{k}_3)\right> \\ && \times 
F_s(\mathbf{k}_3,\mathbf{k}_2;z),
\end{eqnarray}
where $\hat{Q}_s$ is the projection on the subspace orthogonal
to the tagged particle density, $\hat{Q}_s=1-\hat{P}_s$.

Using translational invariance of the sheared suspension and the fact
that in the stationary state the average force acting on the tagged particle
vanishes, we get
\begin{equation}\label{freqmat}
\left<n_s(\mathbf{k}_1)\Omega n_s(\mathbf{k}_3)\right> 
= -D_0 k_1^2 \delta_{\mathbf{k}_1,\mathbf{k}_3}
+ \mathbf{k}_1\cdot\mathbf{\Gamma}\cdot\frac{\partial}{\partial \mathbf{k}_1}
\delta_{\mathbf{k}_1,\mathbf{k}_3},
\end{equation}
where the second term on the right-hand-side (RHS) of Eq. (\ref{freqmat})
describes flow-induced advection.

Furthermore, using the explicit form of the evolution operator we
obtain the following identity:
\begin{eqnarray}\label{memfun}\nonumber
&& \left<n_s(\mathbf{k}_1)\Omega\hat{Q}_s 
\frac{1}{z-\hat{Q}_s\Omega\hat{Q}_s}
\hat{Q}_s\Omega n_s(\mathbf{k}_3)\right> 
\\ \nonumber && = 
\mathbf{k}_1\cdot\left<\mathbf{j}_s(\mathbf{k}_1)  
\frac{1}{z-\hat{Q}_s\Omega\hat{Q}_s} 
\left(2 \mathbf{j}^{\mathrm{eff}}_s(-\mathbf{k}_3) -
\mathbf{j}_s(-\mathbf{k}_3)\right)\right>\cdot\mathbf{k}_3.
\\ &&
\end{eqnarray}
Here $\mathbf{j}_s(\mathbf{k})$ is a projected tagged particle 
current density,
\begin{equation}\label{curr}
\mathbf{j}_s(\mathbf{k})
=
\hat{Q}_s D_0 \left(-i\mathbf{k} + \beta\mathbf{F}_1 + \mathbf{v}_1\right)
e^{-i\mathbf{k}\cdot\mathbf{r}_1},
\end{equation}
and $\mathbf{j}^{\mathrm{eff}}_s(\mathbf{k})$ is a projected, 
effective current density,
\begin{equation}\label{effcurr}
\mathbf{j}^{\mathrm{eff}}_s(\mathbf{k})
=
\hat{Q}_s D_0 \left(-i\mathbf{k} + \beta\mathbf{F}^{\mathrm{eff}}_1 \right)
e^{-i\mathbf{k}\cdot\mathbf{r}_1}.
\end{equation}
In Eq. (\ref{effcurr}) $\mathbf{F}^{\mathrm{eff}}_1$ 
is the effective force acting on the tagged particle
that is defined in terms of the 
stationary, shear-rate dependent probability distribution,
\begin{equation}\label{efffor}
\beta\mathbf{F}^{\mathrm{eff}}_1 = \frac{\partial}{\partial\mathbf{r}_1}
\ln P^{\mathrm{st}}(\mathbf{r}_1, \dots, \mathbf{r}_N).
\end{equation}

Combining Eqs. (\ref{freqmat}-\ref{efffor}), 
in the small wavevector, long time limit we get
\begin{equation}\label{sheardiff}
LT(\dot{F}_s) = \left(-\mathbf{k}_1\cdot\mathbf{D}\cdot\mathbf{k}_1 + 
\mathbf{k}_1\cdot\mathbf{\Gamma}\cdot\frac{\partial}{\partial \mathbf{k}_1}
\right)
F_s(\mathbf{k}_1,\mathbf{k}_2;z),
\end{equation}
where the self-diffusion tensor $\mathbf{D}$ is given by 
the following Green-Kubo-like expression:
\begin{equation}\label{GKdiff}
\mathbf{D} = D_0 - \left(\beta D_0\right)^2 \int_0^{\infty} dt\; 
\left<\mathbf{F}_1 \exp(\Omega t) \left(2 \mathbf{F}^{\mathrm{eff}}_1 
- \mathbf{F}_1\right)\right>.
\end{equation}
Note that in $\mathbf{k}\to 0$ limit
projected dynamics
(\textit{i.e.} $\hat{Q}_s\Omega\hat{Q}_s$) can be replaced by 
real dynamics (\textit{i.e.} $\Omega$) \cite{ED}. 

To get the tagged particle mobility tensor we follow approach 
used by Lekkerkerker and Dhont \cite{LD}.
A calculation along the lines of Sec. III of Ref. \cite{LD} leads
to the following Green-Kubo-like formula for the long-time tagged particle
mobility tensor $\mathbf{\mu}$ (here $\mu_0=\beta D_0$ is the mobility
of an isolated colloidal particle):
\begin{equation}\label{GKmob}
\mathbf{\mu} = \mu_0 - \mu_0^2\beta \int_0^{\infty} dt\; 
\left<\mathbf{F}_1 \exp(\Omega t) \mathbf{F}^{\mathrm{eff}}_1 \right>.
\end{equation}

Comparison of Eqs. (\ref{GKdiff}) and (\ref{GKmob}) 
shows that Einstein's relation 
between self-diffusion and mobility tensors is violated. 
The origin of the violation is the difference
between the force acting on the tagged particle, $\mathbf{F}_1$,
and the effective force, $\mathbf{F}^{\mathrm{eff}}_1$. In the absence of the 
shear flow, 
$\mathbf{F}_1=\mathbf{F}^{\mathrm{eff}}_1$
and the usual Einstein relation follows.
In the presence of the flow, one can follow Ref. \cite{BB} and 
use the transverse components of the self-diffusion and mobility tensors 
to define an effective, shear-rate dependent temperature $T^{\mathrm{eff}}$,
where 
$k_B T^{\mathrm{eff}} = D_{zz}/\mu_{zz}.$
Since we do not have an explicit expression for the effective force 
$\mathbf{F}^{\mathrm{eff}}_1$, neither $D_{zz}$ nor $\mu_{zz}$ can be 
obtained from direct simulational evaluation of its respective
Green-Kubo-like expression. However, if
we obtain $D_{zz}$ from mean-squared displacement 
and measure the force autocorrelation function directly, 
we can obtain the effective temperature:
\begin{equation}\label{GKT}
k_B T^{\mathrm{eff}} = \frac{2D_{zz} k_B T}
{D_0 + D_{zz} - (\beta D_0)^2
\int_0^{\infty}dt\; \left<F_{1z} \exp(\Omega t) F_{1z} \right>}.
\end{equation}
Eq. (\ref{GKT}) shows that it is possible to monitor FDT violation using 
Brownian Dynamics simulations of the stationary, unperturbed state.

Before turning to the derivation of MCT expressions we first
re-write memory function expression (\ref{memfun}). 
We define an irreducible evolution operator $\Omega^{\mathrm{irr}}$,
\begin{equation}
\Omega^{\mathrm{irr}} = -\hat{Q}_s \sum_l \frac{\partial}{\partial\mathbf{r}_l}
\hat{Q}_s \cdot
\left(-\frac{\partial}{\partial\mathbf{r}_l}+\beta\mathbf{F}_l
+\mathbf{v}(\mathbf{r}_l)\right)
\hat{Q}_s,
\end{equation}
and then we use standard projection operator manipulations to obtain
the following identity:
\begin{widetext}
\begin{eqnarray}\label{memfunirr}\nonumber
D_0 \delta_{\mathbf{k}_1,\mathbf{k}_3} 
-\left<\mathbf{j}_s(\mathbf{k}_1)  
\frac{1}{z-\hat{Q}_s\Omega\hat{Q}_s}
\left(2\mathbf{j}^{\mathrm{eff}}_s(-\mathbf{k}_3)  
-\mathbf{j}_s(-\mathbf{k}_3)\right)\right>
= \sum_{\mathbf{k}_4}
\left(\xi_0 \delta_{\mathbf{k}_1,\mathbf{k}_4} + 
\beta\xi_0^2\left<\mathbf{j}_s(\mathbf{k}_1)  
\frac{1}{z-\Omega^{\mathrm{irr}}}
\mathbf{j}^{\mathrm{eff}}_s(-\mathbf{k}_4) \right>\right)^{-1} &&
\\  \times \left(k_B T \delta_{\mathbf{k}_4,\mathbf{k}_3} -
\xi_0 \left<\mathbf{j}_s(\mathbf{k}_4)  
\frac{1}{z-\Omega^{\mathrm{irr}}}
\left(\mathbf{j}^{\mathrm{eff}}_s(-\mathbf{k}_3) -
\mathbf{j}_s(-\mathbf{k}_3)\right)\right>
\right). &&
\end{eqnarray}
\end{widetext}
Note that here $(...)^{-1}$ 
denotes the kernel of the inverse integral operator; also, $\xi_0$
is the friction coefficient of an isolated colloidal particle, 
$\xi_0 = 1/\mu_0$.

Identity (\ref{memfunirr}) allows us to define the tagged particle 
friction tensor 
(the inverse of the first factor at the 
RHS of Eq. (\ref{memfunirr})) and an effective temperature 
(the second factor at the RHS of Eq. (\ref{memfunirr})). It can be shown that 
in the long-time, small wavevector limit, the former reduces to the inverse 
of the mobility tensor, Eq. (\ref{GKmob}).

To derive MCT expressions for the friction tensor and the effective
temperature we follow the standard procedure \cite{GLH,SL}. 
We project the currents
on the part of the joint density of the tagged particle 
and of other particles that is orthogonal to the tagged particle density:
\begin{eqnarray}\label{projection1}
\mathbf{j}_s(-\mathbf{k}) &=& 
\sum_{\mathbf{k}_1,\dots,\mathbf{k}_4} 
n_2(-\mathbf{k}_1,-\mathbf{k}_2) 
g(\mathbf{k}_1,\mathbf{k}_2;\mathbf{k}_3,\mathbf{k}_4) \nonumber \\ &&\times
\left<n_2(\mathbf{k}_3,\mathbf{k}_4)\mathbf{j}_s(-\mathbf{k})\right>,
\end{eqnarray}
\begin{eqnarray}\label{projectione}
\mathbf{j}_s^{\mathrm{eff}}(-\mathbf{k}) &\approx& 
\sum_{\mathbf{k}_1,\dots,\mathbf{k}_4} 
n_2(-\mathbf{k}_1,-\mathbf{k}_2) 
g(\mathbf{k}_1,\mathbf{k}_2;\mathbf{k}_3,\mathbf{k}_4)\nonumber \\ &&\times
\left<n_2(\mathbf{k}_3,\mathbf{k}_4)
\mathbf{j}_s^{\mathrm{eff}}(-\mathbf{k})\right>.
\end{eqnarray}
In Eqs. (\ref{projection1}-\ref{projectione}) 
$n_2(\mathbf{k}_1,\mathbf{k}_2)$ is the part of the 
joint density of the tagged particle 
and of other particles that is orthogonal to the tagged particle density,
$
n_2(\mathbf{k}_1,\mathbf{k}_2) = \hat{Q}_s
\sum_{l>1} e^{-i\mathbf{k}_1\cdot\mathbf{r}_1-i\mathbf{k}_2\cdot\mathbf{r}_l},
$
and $g(\mathbf{k}_1,\mathbf{k}_2;\mathbf{k}_3,\mathbf{k}_4)$
is the inverse of 
$\left<n_2(\mathbf{k}_3,\mathbf{k}_4)n_2(-\mathbf{k}_5,-\mathbf{k}_6)\right>$.
We use factorization approximation for $g$,
$g(\mathbf{k}_1,\mathbf{k}_2;\mathbf{k}_3,\mathbf{k}_4) 
\approx \delta_{\mathbf{k}_1,\mathbf{k}_3} \delta_{\mathbf{k}_2,\mathbf{k}_4} 
\left( N S(\mathbf{k}_2)\right)^{-1},
$
where $S(\mathbf{k}_2)$ is the stationary, shear-rate dependent
structure factor. 
One should note that Eq. (\ref{projection1}) 
is exact for pairwise-additive interactions whereas
Eq. (\ref{projectione}) constitutes an approximation. 

The average in Eq. (\ref{projection1}) 
can be expressed in terms of the direct correlation 
\emph{force} $\mathbf{C}(\mathbf{k})$ \cite{dcforce}, 
\begin{equation}\label{dirforce}
\left<n_2(\mathbf{k}_1,\mathbf{k}_2)\mathbf{j}_s(-\mathbf{k})\right> = 
-i\delta_{\mathbf{k}-\mathbf{k}_1,\mathbf{k}_2}n D_0
\beta\mathbf{C}(\mathbf{k}_2) S(\mathbf{k}_2),
\end{equation}
(here $n$ is the number density, $n=N/V$) 
whereas the average in Eq. (\ref{projectione}) can be expressed in terms the 
non-equilibrium, shear-rate-dependent direct 
correlation function $c(\mathbf{k})=(S(\mathbf{k})-1)/(nS(\mathbf{k}))$, 
\begin{equation}\label{dirfunc}
\left<n_2(\mathbf{k}_1,\mathbf{k}_2)\mathbf{j}_s^{\mathrm{eff}}(-\mathbf{k})
\right> = 
-i\mathbf{k}_2 \delta_{\mathbf{k}-\mathbf{k}_1,\mathbf{k}_2}
n D_0 c(\mathbf{k}_2)S(\mathbf{k}_2).
\end{equation}

Combining Eqs. (\ref{projection1}-\ref{dirfunc}) 
with Eq. (\ref{memfunirr}) we can obtain expressions for the interaction
contributions to the friction tensor and the effective temperature in
terms of integrals involving a four-particle, time-dependent correlation
function. We factorize this
function in terms of the self-intermediate scattering function 
$F_s(\mathbf{k}_1;\mathbf{k}_2;t)$ and
a collective intermediate scattering function
$F(\mathbf{k}_1;\mathbf{k}_2;t)$,
\begin{equation}\label{Fdef}
F(\mathbf{k}_1;\mathbf{k}_2;t) = 
\frac{1}{N}\left<n(\mathbf{k}_1) \exp(\Omega t) n(-\mathbf{k}_2)\right>,
\end{equation}
where 
$n(\mathbf{k}_1)$ is the Fourier transform of the microscopic
density, 
$n(\mathbf{k}_1) = \sum_{l} e^{-i\mathbf{k}_1\cdot\mathbf{r}_l}$.
As a result we obtain the following mode-coupling expressions 
for the interaction contributions to the friction tensor and 
the effective temperature in the long-time (\textit{i.e.} $z\to 0$) limit: 
\begin{widetext}
\begin{equation}\label{friction}
-\beta\xi_0^2\left<\mathbf{j}_s(\mathbf{k}_1)
\left(\Omega^{\mathrm{irr}}\right)^{-1}  
\mathbf{j}^{\mathrm{eff}}_s(-\mathbf{k}_2) \right>\approx
\frac{n}{V}\sum_{\mathbf{k}_3,\dots,\mathbf{k}_6}\int_0^{\infty} dt\; 
\delta_{\mathbf{k}_1-\mathbf{k}_3,\mathbf{k}_4}
\mathbf{C}(\mathbf{k}_4)F_s(\mathbf{k}_3;\mathbf{k}_5;t)
F(\mathbf{k}_4;\mathbf{k}_6;t)c(\mathbf{k}_6)
\mathbf{k}_6\delta_{\mathbf{k}_2-\mathbf{k}_5,\mathbf{k}_6},
\end{equation}
\begin{eqnarray}\label{temp}\nonumber
\xi_0\left<\mathbf{j}_s(\mathbf{k}_1)\left(\Omega^{\mathrm{irr}}\right)^{-1}  
\left(\mathbf{j}^{\mathrm{eff}}_s(-\mathbf{k}_2) 
-\mathbf{j}_1(-\mathbf{k}_2)\right)\right>&\approx &
-\frac{n D_0}{V}
\sum_{\mathbf{k}_3,\dots,\mathbf{k}_6}
\int_0^{\infty} dt\; \delta_{\mathbf{k}_1-\mathbf{k}_3,\mathbf{k}_4}
\mathbf{C}(\mathbf{k}_4)F_s(\mathbf{k}_3;\mathbf{k}_5;t)
F(\mathbf{k}_4;\mathbf{k}_6;t)\\ && \times
\left(c(\mathbf{k}_6)\mathbf{k}_6
-\beta\mathbf{C}(\mathbf{k}_6)\right)
\delta_{\mathbf{k}_2-\mathbf{k}_5,\mathbf{k}_6}.
\end{eqnarray}
\end{widetext}

Expression (\ref{friction}) differs from one derived before
\cite{IR}: one of the vertices in Eq. (\ref{friction}) 
involves the direct correlation force,
whereas the other involves the non-equilibrium direct correlation function. 
In the expression obtained in Ref. \cite{IR} 
both vertices were identical and given by the equilibrium 
direct correlation function. One should note
that in equilibrium $\mathbf{C}^{eq}(\mathbf{k}) = k_B T \mathbf{k}
c^{eq}(k)$; hence the previous work implicitly 
used equilibrium approximation for the vertices.

Expression (\ref{temp}) shows that FDT violation is associated
with the non-equilibrium character of the stationary, sheared state.
In particular, if equilibrium approximation for the vertices is used,
no FDT violation is obtained.

One of the most interesting simulational findings is
that below the MCT transition temperature FDT violation persists
in the limit of the vanishing shear rate \cite{BB}. This can be qualitatively
understood on the basis of Eq. (\ref{temp}): in the $\dot{\gamma}\to 0$
limit the second vertex in (\ref{temp}) vanishes; however, 
in the same limit characteristic relaxation
times of the self and collective intermediate scattering functions 
diverge. Thus, it is possible that the expression 
(\ref{temp}) reaches a finite limit as $\dot{\gamma}\to 0$.
In order to prove that this indeed happens one needs 
to calculate the second vertex in Eq. (\ref{temp}). 
To this end, it might be possible to 
use an approach proposed by Fuchs and Cates \cite{FC}:  
using mode-coupling theory to calculate steady state properties by 
starting from the equilibrium state and 
considering transient dynamics.

Another very interesting result of Ref. \cite{BB} is that 
in the $\dot{\gamma}\to 0$ limit the same
effective temperature is obtained for different wavevectors and
even for different observables. Expression (\ref{temp}),
in principle, allows one to verify this fact theoretically. In particular
one could check whether \emph{tensorial} quantity (\ref{temp}) reduces
to a scalar one in the $\dot{\gamma}\to 0$ limit. Again,
in order to investigate this, the second vertex is needed.

To summarize, we have derived Green-Kubo-like formulae for 
the self-diffusion and mobility coefficients, and mode-coupling
expressions for the friction tensor and the effective
temperature. The numerical analysis of these expressions is left for
future work. 

The author would like to thank Matthias Fuchs, David Reichman 
and Kunimasa Miyazaki for stimulating discussions; support by
NSF Grant No. CHE-0111152 is gratefully acknowledged.


\begin{thebibliography}{99}
\bibitem{commentHI2} Following prior works on colloids under shear 
\cite{FC,MR,IR,SBL}, hydrodynamic interactions are neglected.
\end{thebibliography}
\end{document}